\documentclass[12pt]{article}
\usepackage{amssymb,amsmath,amsthm,graphicx,ulem}

\pdfoutput=1
\usepackage{hyperref}
\usepackage{graphicx,subfigure}
\usepackage{epsfig}
\usepackage{amsmath}
\usepackage{amsfonts}
\usepackage{amssymb}
\usepackage[usenames]{color}
\usepackage[letterpaper,left=2.cm,right=2.cm,top=2.5cm,bottom=2.5cm]{geometry}

\newcommand{\beq}{\begin{equation}}
\newcommand{\eeq}{\end{equation}}
\newcommand{\be}{\begin{equation}}
\newcommand{\ee}{\end{equation}}
\newcommand{\beqa}{\begin{eqnarray}}
\newcommand{\eeqa}{\end{eqnarray}}
\newcommand{\beqar}{\begin{eqnarray*}}
\newcommand{\eeqar}{\end{eqnarray*}}
\newcommand{\bea}{\begin{eqnarray}}
\newcommand{\eea}{\end{eqnarray}}

\numberwithin{equation}{section}

\newcommand{\nn}\nonumber
\newcommand{\eqn}[1]{(\ref{#1})}
\numberwithin{equation}{section}

\begin{document}

\allowdisplaybreaks

\normalem

\title{General Relativity and the Cuprates}

\author{Gary T. Horowitz and
Jorge E. Santos\\ 
\\ \\
 Department of Physics, UCSB, Santa Barbara, CA 93106, USA \\ 
 \\ 
 \small{gary@physics.ucsb.edu, jss55@physics.ucsb.edu}}

 \date{}

\maketitle

\begin{abstract}
\noindent  We add a periodic potential to the simplest gravitational model of a  superconductor and compute the optical conductivity. In addition to a superfluid component, we find a normal component that has Drude behavior at low frequency followed by a power law fall-off. Both the exponent and coefficient of the power law are temperature independent and agree with earlier results computed above $T_c$. These results are in striking agreement with measurements on some cuprates. We also find  a gap $\Delta = 4.0\ T_c$, a rapidly decreasing scattering rate, and ``missing spectral weight" at low frequency, all of  which also agree with experiments.
 \end{abstract}

\newpage

%%%%%%%%%%%%%%%%%%%%%%%%%%%%%%%%%%%%%%%%%%%%%%%%%%%%%%%%%%%%%%%%%%%%%%%%%%%
%%%%%%%%%%%%%%%%%%%%%%%%%%%%%%%%%%%%%%%%%%%%%%%%%%%%%%%%%%%%%%%%%%%%%%%%%%%

\tableofcontents
\baselineskip16pt

\section{Introduction}
%%%%%%%%%%%%%%%%%%%%%%%%%%%%%%%%%%%%%%%%%%%%%%%%%%%%%%%%%%%%%%%%%%%%%%%%%%%

Gauge/gravity duality has provided a powerful new tool to analyze certain condensed matter systems. In particular, it can be used to calculate transport properties of strongly correlated systems at finite temperature. Remarkably, this is achieved by mapping the problem to a gravitational problem in one higher dimension. Since the system of interest lives in one dimension less than the gravitational problem being solved, this approach is often called holographic.

The optical conductivity in a simple holographic model of a $2+1$ dimensional conductor was recently studied including the effects of a lattice  \cite{Horowitz:2012ky}. Earlier studies assumed translation invariance which implied momentum conservation. In that case, the charged particles cannot dissipate their momentum so the real part of the optical conductivity always contains a delta function at zero frequency reflecting infinite DC conductivity. With the lattice included, the delta function is resolved.  It was found that at low frequency the conductivity follows the simple Drude form, but at intermediate frequency, it follows a power law
\be |\sigma(\omega)| = \frac{B}{\omega^{2/3}}+C\label{power}\ee
The exponent $-2/3$ is surprisingly robust, and is independent of the all the parameters in the  model including lattice spacing, lattice amplitude, and temperature. 
Strikingly, a power-law at mid-infrared frequencies with exactly this exponent is seen in measurements of the normal phase of bismuth-based  cuprates   \cite{vandermarel,marel2}. (See also \cite{azrak} for earlier experimental results and \cite{anderson,kato,norman} for attempts to explain this scaling). Also in agreement with the measurements is the fact that the coefficient $B$ is temperature independent, although  the cuprates do not appear to have the off-set $C$.

The origin of \eqn{power} from the gravitational side is still not understood. It has been shown that a similar power-law is seen in the thermoelectric conductivity and for a $3+1$ dimensional conductor (although the exponents differ) \cite{Horowitz:2012gs}.  In a recent paper, Vegh \cite{Vegh:2013sk} studied  the optical conductivity in a holographic model in which momentum conservation was broken by a graviton mass term. He finds a similar power-law scaling of the  conductivity (including the constant off-set), but the  exponent is not fixed. It depends on the mass of the  graviton.

The goal of this paper is to extend these results to the superconducting regime. We add a lattice in the form of a periodic potential to the simplest (original) holographic superconductor \cite{Hartnoll:2008vx,Hartnoll:2008kx} and compute the optical conductivity. We find that below $T_c$, in addition to the superfluid, there is a normal fluid component  with the following properties

\begin{itemize}

\item It has simple Drude behavior at low frequency
\be\label{drude}
\sigma(\omega) = {\rho_n \tau  \over 1-i \omega \tau}
\ee
where the normal component density $\rho_n$ and  relaxation time $\tau$ are temperature dependent (but frequency independent).

\item The scattering rate $1/\tau$ drops rapidly below $T_c$.

\item At intermediate frequency, the normal component again satisfies the power law \eqn{power} with the same coefficient $B$ that was seen above $T_c$.

\item There is evidence for a superconducting gap of size $\Delta = 4.0\ T_c$.

\item Despite this, in the limit of low temperature, $\rho_n$ does not vanish, indicating the presence of uncondensed spectral weight.

\item The Ferrel-Grover-Tinkham sum rule relating the superfluid density to the decrease in  Re$[\sigma(\omega)]$ is satisfied, but only if large frequencies of order the chemical potential are included. If one considers only the Drude peak and power law region, there is ``missing spectral weight".

\end{itemize}

As we will discuss, these are all observed properties of the bismuth-based  cuprates. In particular, experiments have shown that the  power-law is unchanged when the temperature drops below the critical temperature and the material becomes superconducting \cite{hwang}.

\section{Gravitational model}

We will work with the simplest holographic superconductor which requires gravity coupled to a Maxwell field and charged scalar $\Phi$. The action is 
\begin{equation}
S= \frac{1}{16 \pi G_N}\int d^4 x\,\sqrt{-g}\left[R+\frac{6}{L^2}-\frac{1}{2}F_{ab}F^{ab}-2|(\nabla - i\,e\,A) \Phi|^2  +\frac{4|\Phi|^2}{L^2}\right],
\label{eq:action}
\end{equation}
where  $L$ is the AdS length scale and $F = dA$. The scalar mass, $m^2 = -2/L^2$, is chosen  since for this choice, the asymptotic behavior of $\Phi$ is simple. From here on we work in units in which $L=1$. If the metric asymptotically takes the usual form
\be ds^2 = \frac{-dt^2 + dx^2 + dy^2 + dz^2}{z^2}
\ee
then
\be
\Phi = z\phi_1 + z^2\phi_2 + {\cal O}(z^3)\,.
\label{eq:asymptoticPhi}\ee
In the standard interpretation, $\Phi$ is dual to a dimension two charged operator $O$ in the dual theory with source $\phi_1$ and expectation value $\phi_2$. Since we want this condensate to turn on without being sourced, we set $\phi_1 =0$.

As Gubser first suggested \cite{Gubser:2008px}, the above action has the property that (electrically) charged black holes become unstable at low temperatures to developing scalar hair. The reason is essentially that the effective mass of the scalar gets a contribution $e^2 A_t^2 g^{tt} < 0$   from its coupling the Maxwell field which causes the $\Phi =0$ solution to become unstable. It was shown in \cite{Hartnoll:2008vx,Hartnoll:2008kx} that this is precisely the gravitational dual of a conductor/superconductor phase transition.

The vector potential $A_t$ 
must vanish at the horizon and is asymptotically
\be\label{asympAt}
A_t = \mu - \rho z + O(z^2)
\ee
where $\mu$ is the chemical potential and $\rho(x)$ is the charge density.  For these electrically charged solutions, the field equations and boundary conditions require the phase of $\Phi$ to be constant. We will set it to zero and treat $\Phi$ as a real field.
We introduce the lattice by requiring that the chemical potential be a periodic function of $x$:
\be  
\mu(x) =\bar{\mu} \left[1 + A_0\cos(k_0x)\right]\,.
\label{wavy}\ee
The lattice is only introduced in one direction  for computational convenience, and we will compute the conductivity only in the direction of the lattice. This ``ionic" lattice has been discussed earlier (see, e.g., 
\cite{Hartnoll:2012rj,Maeda:2011pk,Liu:2012tr,Flauger:2010tv,Hutasoit:2012ib}) but almost always treated perturbatively. In \cite{Horowitz:2012gs} it was treated  exactly in the theory \eqn{eq:action} without the charged scalar field.

Our model has several parameters. In principle one could vary the mass and charge of the scalar field $\Phi$ in the action. However, we have already fixed the mass for convenience. The boundary condition (\ref{wavy}) depends on three parameters: the mean chemical potential $\bar \mu$, the lattice wavenumber $k_0$ and the lattice amplitude $A_0$. In addition, the solution will depend on a temperature $T$. Due to a scaling symmetry, physics depends on only three dimensionless quantities which can be taken to be $A_0$, $k_0/\bar{\mu}$,  and $T/\bar{\mu}$.

For definiteness, we will choose $k_0/\bar{\mu} = 2$ for the results presented later in this paper. Although the physics is scale invariant, when doing calculations, we will also set $\bar\mu =1$. Our choice of $k_0$ is very close to the one which yields a low temperature DC resistivity which is linear in $T$ in the absence of superconductivity \cite{Hartnoll:2012rj,Horowitz:2012ky}. Even though $k_0$ corresponds to an unrealistically large lattice spacing,  the nonlinearity of Einstein's equation generates structure on much smaller scales. 

\section{Backgrounds}
To numerically  construct the gravitational dual of the normal phase we employ the ansatz used in \cite{Horowitz:2012ky,Horowitz:2012gs}:
\begin{equation}
ds^2 = \frac{1}{z^2}\left[-H_1\,G(z)\,(1-z)dt^2+\frac{H_2\,dz^2}{(1-z)\,G(z)}+S_1\,(dx+F\,dz)^2+S_2\,dy^2\right],\quad A = \psi\, dt,\quad \Phi=0,
\label{eq:ansatz_normal_phase}
\end{equation}
and $G(z) = 1+z+z^2-\mu_1^2\,z^3/2$. Here $G(z)$ controls the black hole temperature given by $T \equiv G(1)/4\pi = (6-\mu_1^2)/8\pi$ and $H_{1,2}$,  $S_{1,2}$, $F$ and $\psi$ are six functions of $x$ and $z$, that we determine using the numerical methods described in \cite{Horowitz:2012ky}, which were first introduced in \cite{Headrick:2009pv} and studied in great detail in \cite{Figueras:2011va}.

When $A_0 = 0$ the problem reduces to the original holographic superconductor \cite{Hartnoll:2008kx}. It was shown there that there is a critical value of $T/\mu$ such that above this value, the scalar field vanishes and the solution is the simple planar Reissner-Nordstr\"om AdS metric. Below this value, the Reissner-Nordstr\"om solution becomes unstable to developing scalar hair. 

We now increase $A_0$ and solve the Einstein-Maxwell equations to find rippled versions of  the Reissner-Nordstr\"om AdS metric. This is exactly the same as what was done for the ionic lattice in \cite{Horowitz:2012gs}.  To see when these solutions become unstable to forming scalar hair, we look for a static normalizable mode of the scalar field.  It is intuitively clear that if one increases the charge $e$ on the scalar field, it becomes easier for this field to condense and the critical temperature becomes higher. To find the critical temperature for a given charge it is convenient to turn the problem around.  For each $A_0$ and any temperature $T$, we find the value of the charge $e$ such that $T$ is the critical temperature for a field with charge $e$. At the onset of the instability, $\Phi$ is a static normalizable mode of the charged scalar field and can be treated perturbatively:
\begin{equation}
(\nabla^a- i\,e\,A^a)(\nabla_a- i\,e\,A_a)\Phi+2\,\Phi = 0,
\end{equation}
with $\Phi$ being a function of $x$ and $z$ only. Also, the connections $\nabla$ and $A$ are evaluated on the rippled Reissner-Nordstr\"om AdS black holes. Finally, one can recast the former equation to take the following appealing form
\begin{equation}
-\nabla^a\nabla_a\Phi-2\,\Phi =e^2\,(-A_a A^a)\,\Phi,
\end{equation}
which one recognizes as a positive selfadjoint eigenvalue problem for $e^2$. At the boundary we demand $\Phi$ to decay as in Eq.~(\ref{eq:asymptoticPhi}) and at the horizon we demand regularity.

\begin{figure}
\centerline{
\includegraphics[width=.5\textwidth]{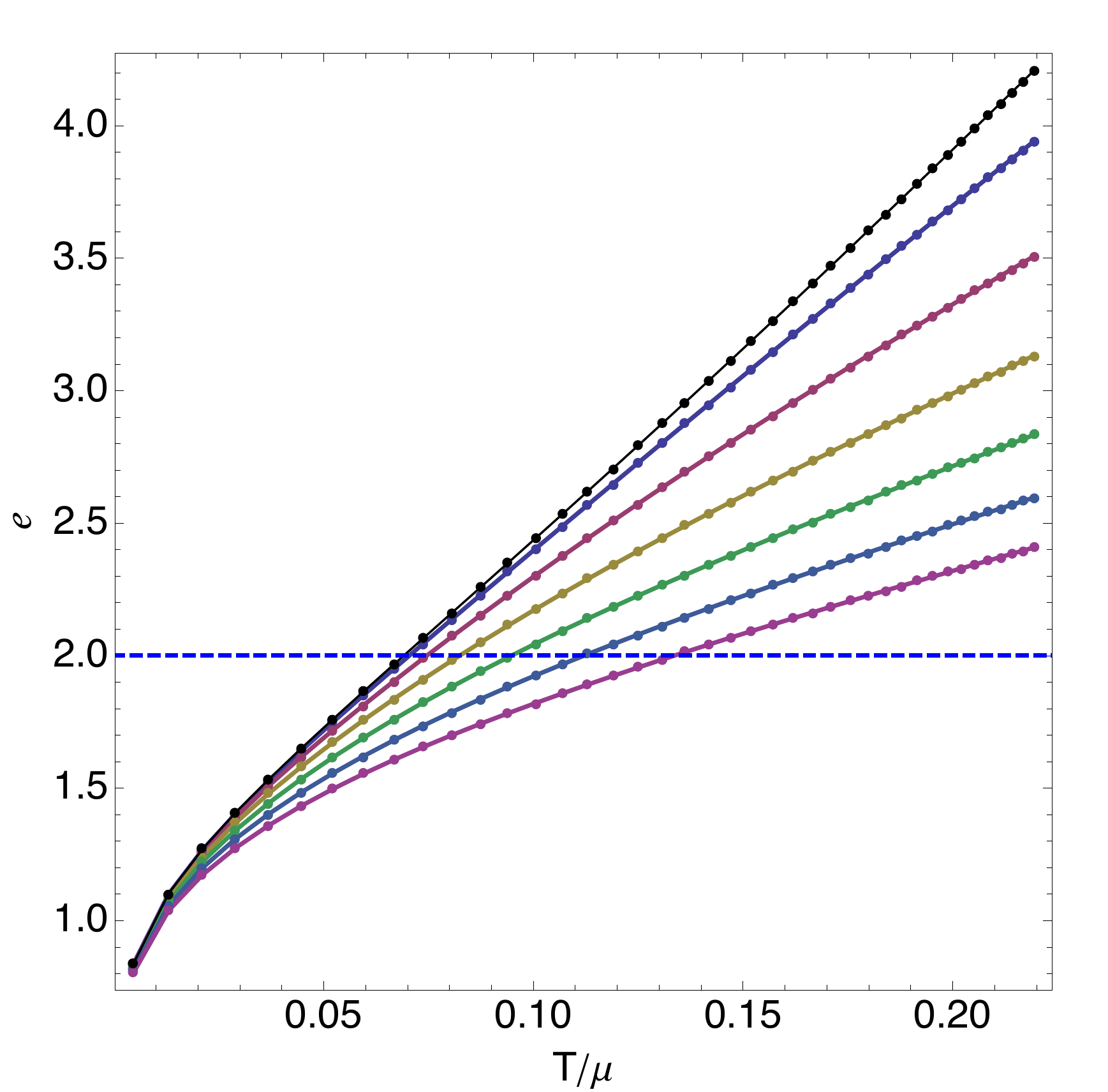}
}
\caption{For each $T/\mu$ we plot the charge of the scalar field for which $T$ would be the critical temperature. This is repeated for several values of the amplitude of the lattice. From the top down the lines represent: $A_0 = 0.0, 0.4, 0.8, 1.2, 1.6, 2.0, 2.4$. Setting $ e =2$ we read off the critical temperatures used in this paper.}
\label{fig:charge}
\end{figure}

The results are shown in Fig. \ref{fig:charge}.  There are seven curves on this plot denoting seven different values of the lattice amplitude $A_0$ between $0$ and $2.4$. Each curve shows the expected rise in critical charge with temperature (or critical temperature with charge). Comparing the different curves for a given charge, we see that increasing the lattice amplitude also increases the critical temperature. This was first noticed in \cite{Ganguli:2012up}  when the lattice was treated perturbatively. It can be understood as follows. As we said, the critical temperature of a uniform holographic superconductor is proportional to $\mu$. By making $\mu$ vary periodically, one raises the maximum value of $\mu$ which induces the scalar field to condense at a higher temperature.  It then leaks into the regions where $\mu$ is smaller. This effect was seen in models of a holographic Josephson junction \cite{Horowitz:2011dz}.  We will set $e = 2$ since this is a natural value for a superconducting condensate.

Having found the critical temperature, one can next solve the coupled Einstein-Maxwell-scalar equations to find the solutions for  $T < T_c$. In order to find the rippled superconducting phase we have to change the ansatz (\ref{eq:ansatz_normal_phase}). The reason being that Eq.~(\ref{eq:ansatz_normal_phase}) is adapted\footnote{By adapted here we mean that the $T\to0$ limit can be achieved without introducing large gradients on the functions $S_{1,2}$.} to probe geometries for which the entropy is nonzero as $T\to0$. The homogeneous holographic superconductors are known not to have this feature. In particular, their entropy decreases towards zero as the temperature is lowered, thus obeying to the third law of thermodynamics. One ansatz that is adapted to this property is given by
\begin{equation}
ds^2 = \frac{1}{z^2}\left\{-H_1\,y_+^2\,(1-z)dt^2+\frac{H_2\,dz^2}{(1-z)}+y_+^2\left[S_1\,(dx+F\,dz)^2+S_2\,dy^2\right]\right\},\quad A = \psi\, dt
\label{eq:ansatz_superconducting_phase}
\end{equation}
where $H_{1,2}$,  $S_{1,2}$, $F$, $\psi$ and $\Phi$ are seven functions of $x$ and $z$ to be determined using the numerical methods introduced in \cite{Horowitz:2012ky}. Here, $y_+$ parametrizes the black hole temperature as $T = y_+/4 \pi$ and, in the homogenous case, is the black hole horizon size measured in units of the AdS radius.

\begin{figure}
\centerline{
\includegraphics[width=.5\textwidth]{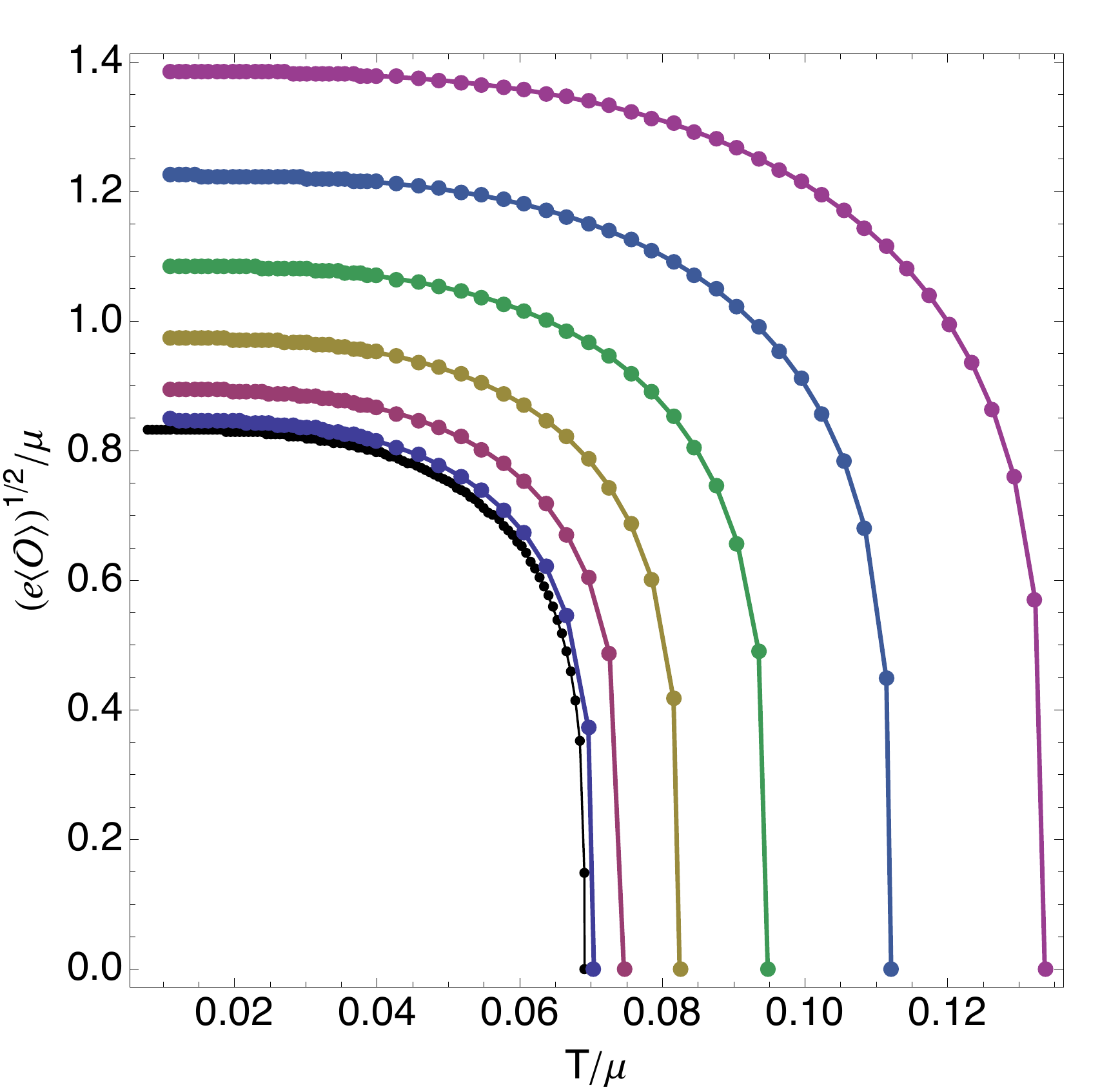}
}
\caption{The mean value of the condensate as a function of temperature for $e = 2$ and various values of the lattice amplitude. From the inner to outer curves: $A_0 = 0.0, 0.4, 0.8, 1.2, 1.6, 2.0, 2.4$. The colors agree with Fig. \ref{fig:charge}.}
\label{fig:condensate}
\end{figure}

\begin{figure}
\centerline{
\includegraphics[width=.5\textwidth]{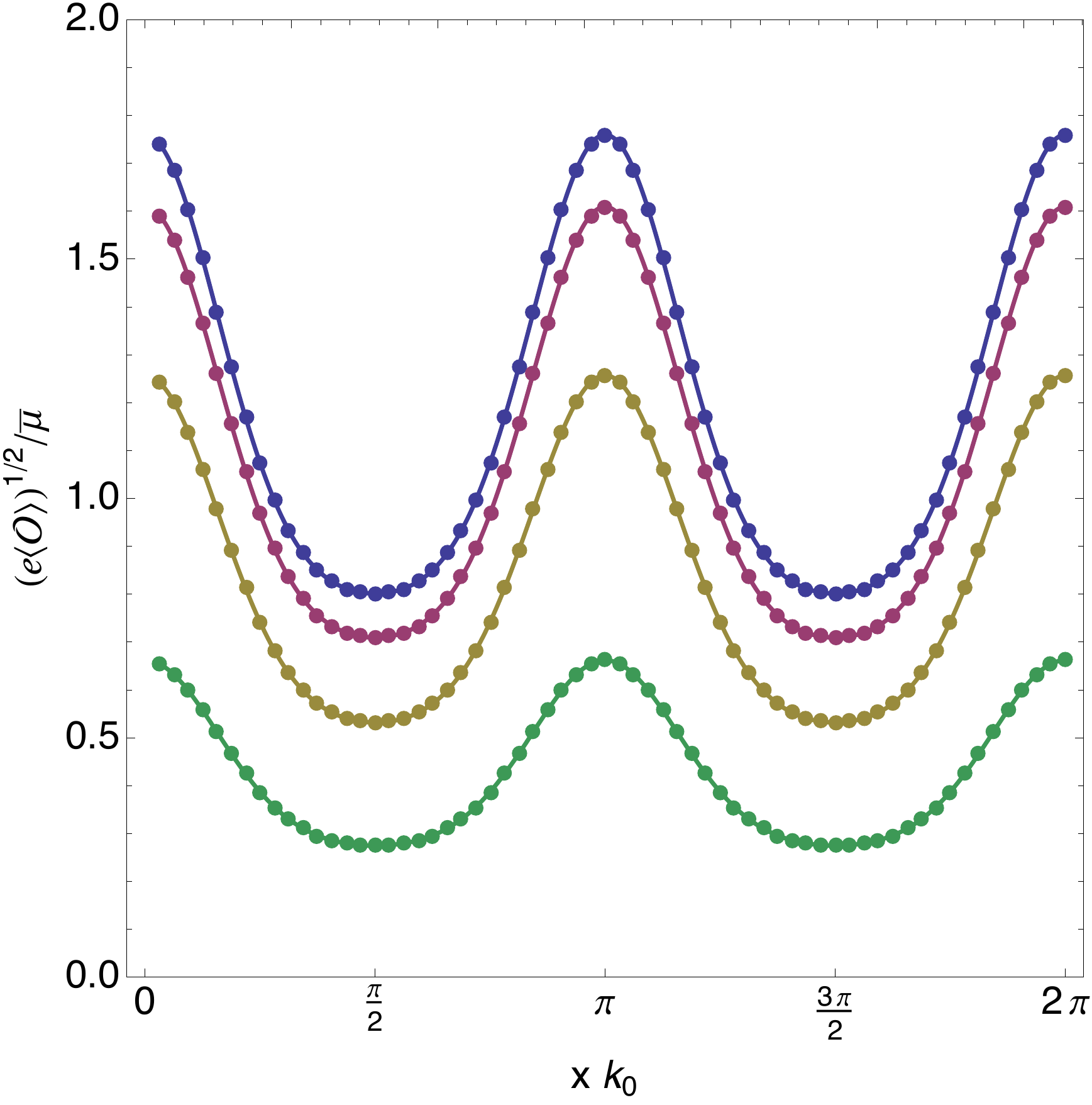}
}
\caption{The condensate as a function of $x$ for $e=2$, $A_0 = 2$,  and $T/T_c = .2, .7, .9, .99$ from top down. Note that by $T/T_c = .7$ the condensate has almost reached its low temperature limit.}
\label{fig:condensate2}
\end{figure}

The result are hairy, rippled, charged black holes. From the asymptotic form of the scalar field one reads off the expectation value of the charged condensate in the dual theory. The condensate is a function of $x$, oscillating about a mean value. In Fig. \ref{fig:condensate} we plot this mean value as a function of temperature for various values of the lattice amplitude. In each case, the condensate follows a familiar form, rising rapidly as $T$ drops below $T_c$ and then saturating at low temperature. It is clear from Fig. \ref{fig:condensate} that increasing the lattice amplitude increases the low temperature mean value of the condensate. In Fig. \ref{fig:condensate2} we show the variation in the condensate by plotting $\langle O(x) \rangle$ for $A_0 = 2$ and various temperatures. The variation is roughly $50\%$ of the mean and the condensate remains positive always. 

The transition at $T=T_c$ which turns on the scalar condensate corresponds to a continuous phase transition. One can show this without explicitly computing the free energy as follows. Consider first the homogeneous case with constant chemical potential $\mu$, and compactify $x$ and $y$. The free energy in the grand canonical ensemble is $F = E - TS - \mu Q$, so its variation is 
\be
\delta F = \delta E -T \delta S - \mu \delta Q - S \delta T - Q \delta\mu
\ee
But the first law says that $\delta E =T \delta S + \mu \delta Q $, so at fixed $\mu$,
\be\label{secondorder}
\frac{dF}{dT} = -S
\ee
Since the branch of solutions with scalar hair joins the branch of solutions without hair at $T=T_c$, their entropies must agree, and hence $dF/dT$ is continuous.

This argument generalizes to the case of a position dependent chemical potential.
The $\mu Q$ term in the free energy is replaced by $\int \mu(x) \rho(x) dx dy$ and the $\mu \delta Q$ term in the first law is replaced by $\int \mu(x) \delta \rho(x) dx dy$ (see Appendix).  Since we are keeping $\mu(x)$ fixed as we change the temperature, we recover \eqn{secondorder}.  The two branches of solutions with ripples still join at $T=T_c$, so the phase transition is again continuous.

Since we have only introduced the lattice in one direction, one could think of our solutions as representing a striped superconductor. However we will not pursue that interpretation here.\footnote{It is not a gravitational dual of the novel striped superconductor introduced in \cite{Berg} since that required a condensate whose average value was zero.}

In the homogeneous case, the zero temperature limit of these hairy black holes is known to take the form \cite{Horowitz:2009ij}
\be\label{nhmetric}
ds^2 = r^2(-dt^2  + dx_idx^i) + {dr^2\over g_0 r^2(-\log r)}
\ee
and $\Phi =2(-\log r)^{1/2}$ near $r=0$. This metric has a null singularity at $r=0$. The scalar field on the horizon of our solutions is becoming more homogenous as $T \rightarrow 0$, and at low temperatures, the entropy scales like $S \propto T^{2.4}$ independent of the lattice amplitude. However, the coefficient in the entropy formula does depend on the lattice amplitude, so it is not clear if \eqn{nhmetric} applies to the $T=0$ limit of our rippled solutions.

The solutions we have discussed so far are not the only static, rippled, hairy black hole solutions. There are radial excitations where the scalar field has nodes in the radial direction. There are also excitations in the direction of the lattice which can be constructed as follows. Given a solution $\Phi$ with period $\lambda_0 =2\pi/k_0$, $-\Phi$ is clearly also a solution. One can construct a solution in which $\Phi$ changes sign from one region of size $\lambda_0$ to the next, by simply starting with a seed solution in Newton's iterative method that changes sign. Given this, one can construct solutions with period $(n+m)\lambda_0$  in which $\Phi$ changes sign on $m$ of the regions. This would clearly result in a smaller mean value for the condensate. We expect that these other solutions all have higher free energy than the ones we study, in which $\Phi$ remains positive everywhere.

%%%%%%%%%%%%%%%%

\section{Conductivity}

To compute the optical conductivity in the direction of our lattice  we  introduce a perturbation with harmonic time dependence and  fix the usual boundary condition on $\delta A_x$: 
\be
 \delta A_x \rightarrow \frac{E}{i\omega} + J_x(x, \omega) z + O(z^2)\,.
 \label{Abdycond}
\ee
This corresponds to adding a homogeneous electric field $E_x = E e^{-i\omega t}$  on the boundary. This perturbation induces perturbations in most metric components as well as other components of the vector potential and scalar.
Solving these linear equations with suitable boundary conditions allows us to read off the current $J_x(x, \omega)$ which determines the conductivity via $\widehat{\sigma}(\omega,x) = J_x(x, \omega)/E$.  In contrast with the ionic lattice of \cite{Horowitz:2012gs}, there are now a total of twelve linear functions to be determined. In addition to the perturbations present in the ionic case, we also have two more perturbations corresponding to the real and imaginary parts of the charged scalar field. We could have worked with $U(1)$ gauge invariant variables, such as in \cite{Horowitz:2011dz}, however here we follow a standard gauge fixing procedure. For the metric and gauge field perturbations we use the de Donder and Lorentz gauges, respectively.
%We can express the conductivity in a manifestly gauge invariant form as
% \begin{equation}
%\tilde{\sigma}(\omega,x)\equiv\lim_{z\to 0}\frac{\delta F_{zx}(x,z)}{\delta F_{xt}(x,z)}.
%\label{eq:opticalderiv}
%\end{equation}
 Since we  impose a homogeneous electric field, we are interested in the homogeneous part of the conductivity, $\sigma(\omega)$, that can be retrieved from the full conductivity via Fourier mode decomposition:
 \begin{equation}
 \widehat{\sigma}(\omega,x) = \sigma(\omega)+\sum_{\begin{subarray}{c} n=-\infty\\ n\neq 0 \end{subarray}}^{+\infty} \sigma_n(\omega)e^{i\,n\,k_0\, x}.
 \end{equation}
 
 \begin{figure}
\centerline{
\includegraphics[width=.8\textwidth]{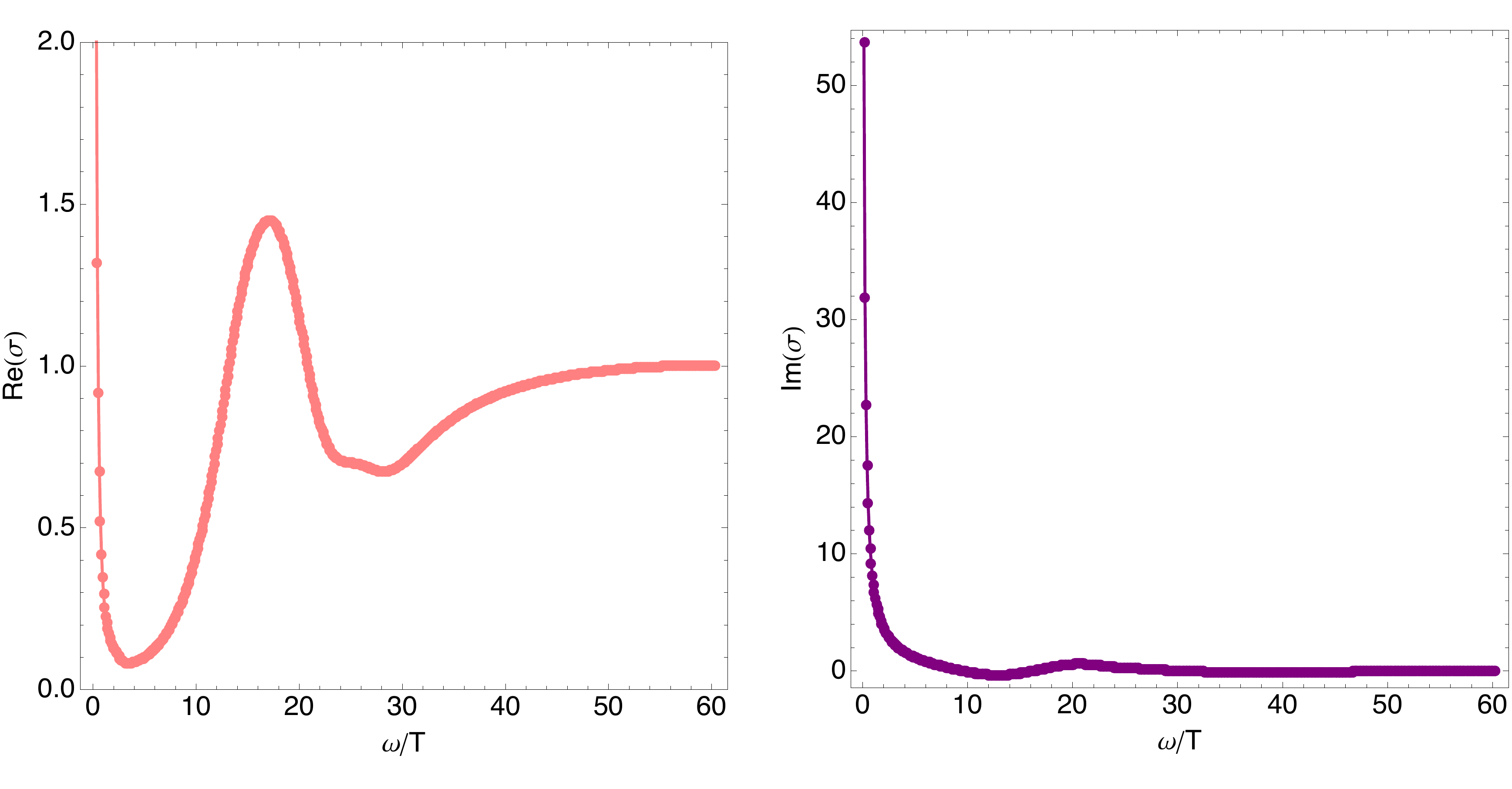}
}
\caption{The real and imaginary parts of the conductivity for $A_0 =2, k_0 = 2$, and $T/T_c = .71$. }
\label{fig:longconductivity}
\end{figure}

In this section, we will fix the lattice amplitude to be $A_0 = 2$, which corresponds to a critical temperature $T_c =.11\bar\mu$.

The real and imaginary parts of the conductivity for $T/T_c = .71$ are shown in Fig. \ref{fig:longconductivity}.  The first thing to note is that there is a pole in the imaginary part of the conductivity.  This implies (from the Kramers-Kronig relation) that there is a delta function at $\omega = 0$ in the real part confirming that this is  a superconductor. So, unlike the normal phase,  the holographic lattice does not remove the delta function in the superconducting regime\footnote{This was noticed earlier in a perturbative treatment of the lattice \cite{Hutasoit:2012ib,Iizuka:2012dk}.}.  The coefficient of the pole is the superfluid density. The bump in the real part of $\sigma$ at $\omega/T \approx 20$ is a resonance which is analogous to the one studied in \cite{Horowitz:2012gs} coming from a quasinormal mode of the black hole. We will return to this later, but for now, the most interesting feature of Re($\sigma$)  is the rise at low frequency.  
This shows that there is a normal component to the conductivity even in the superconducting regime. Thus, our holographic superconductor resembles a two fluid model with a normal component as well as a superfluid component.

\subsection{Low frequency region}

\begin{figure}
\centerline{
\includegraphics[width=.8\textwidth]{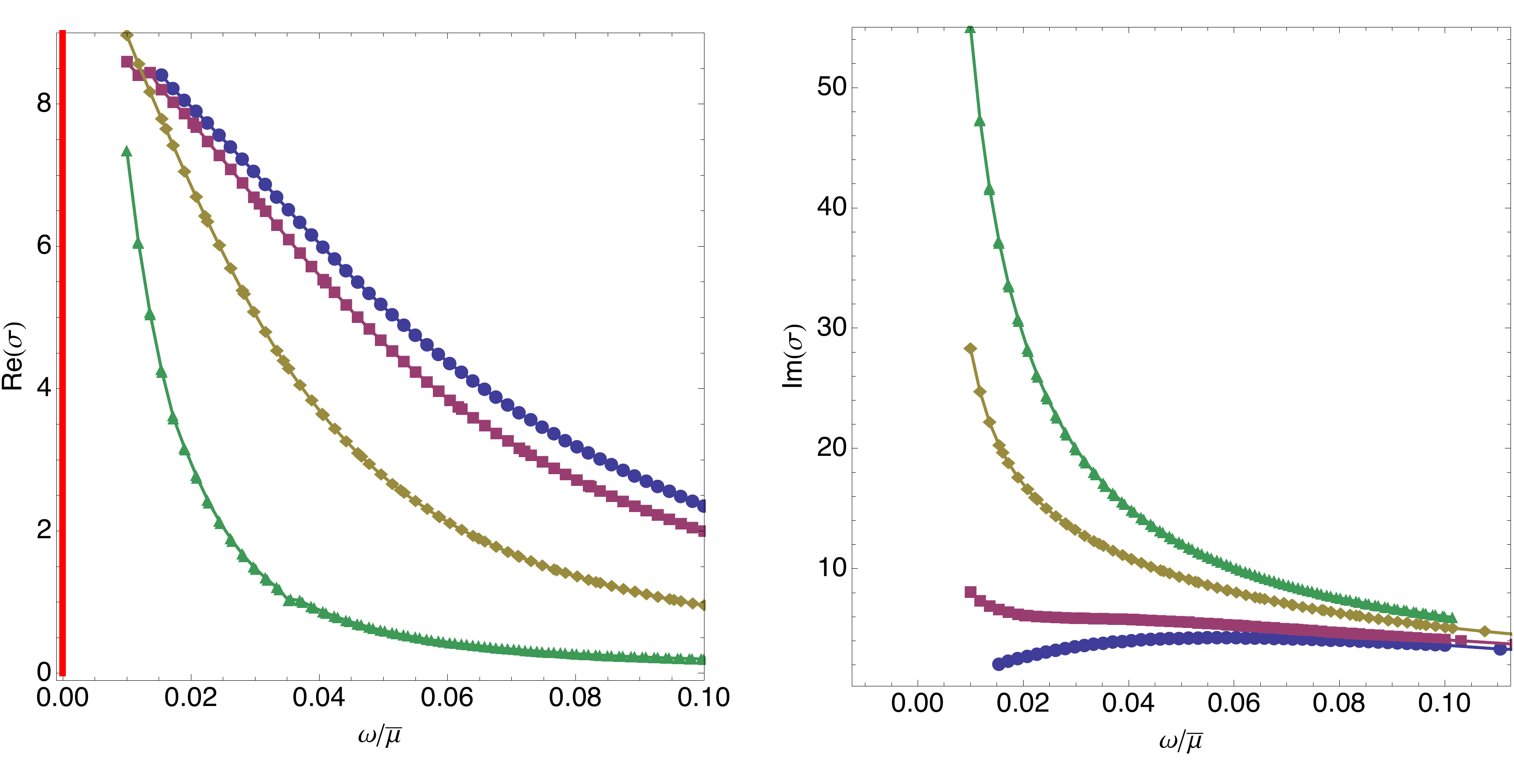}
}
\caption{The low frequency part of the conductivity for $T/T_c = 1.0$ (blue circles), .97 (red squares), .86 (yellow diamonds),  and .70 (green triangles). The vertical red line  in Re$(\sigma)$ denotes the zero frequency delta function.}
\label{fig:conductivity}
\end{figure}

Let us now focus on this low frequency region. The real and imaginary parts of the conductivity for various temperatures are shown in Fig. \ref{fig:conductivity}.  There are four curves ranging from $T/T_c = 1$ down to $T/T_c = .7$. It should be emphasized that the curve with $T/T_c = 1$ was done with a completely different numerical code using a different metric ansatz and solving for one fewer unknown function (no scalar). The continuity of the results is a good check on the numerical accuracy. It is  clear from Fig. \ref{fig:conductivity} that as we lower the temperature, the normal component is decreasing, and the pole in the imaginary part is increasing, showing an increase in the superconducting component.

A closer examination of the normal component of the conductivity when $T<T_c$ shows that it behaves very much like the conductivity above the critical temperature. In fact, at low frequency both the real and imaginary parts of the conductivity are very well fit by simply adding a pole to the  Drude formula:
\be\label{drudesuper}
\sigma(\omega) = i{\rho_s\over \omega} + {\rho_n \tau  \over 1-i \omega \tau}
\ee
where $\rho_s$ is the superfluid density, $\rho_n$ is the normal fluid density, and $\tau $ is the relaxation time. The three parameters $\rho_s$, $\rho_n$ and $\tau$ are temperature dependent but frequency independent.  On the right hand side of Fig. \ref{fig:conductivity} one can clearly see the Drude behavior in the bottom curve which is still in the normal phase. As the temperature is lowered, the pole grows rapidly and soon swamps the Drude behavior of the normal component. 

 \begin{figure}
\centerline{
\includegraphics[width=.8\textwidth]{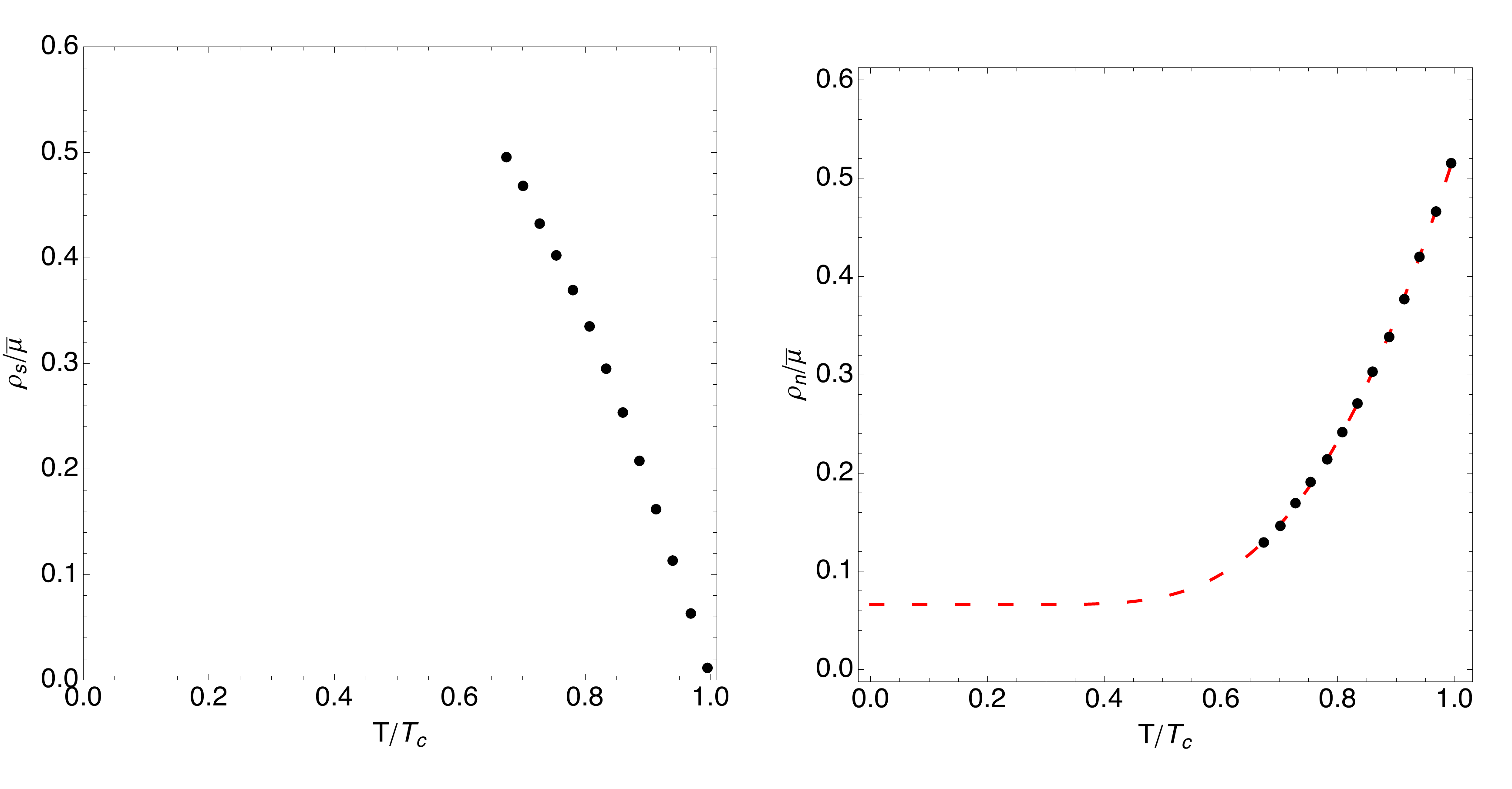}
}
\caption{The density of the superfluid and normal components as a function of temperature, extracted from the fit to \eqn{drudesuper}. The dashed red line on the right is a fit to \eqn{rhon}.}
\label{fig:densities}
\end{figure}

The temperature dependence of $\rho_s$ and $\rho_n$ are shown in Fig.~\ref{fig:densities}. On the left one sees that $\rho_s$ rises rapidly as $T$ drops below $T_c$, and on the right, one sees that $\rho_n$ drops rapidly. The red dashed line is a fit to 
\be\label{rhon}
\rho_n(T) = a + b\,e^{-\Delta /T} \quad {\rm with} \quad \Delta = 4.0 \ T_c\,.
\ee
 This is similar to a BCS superconductor in which the normal component consists of thermally excited quasiparticles with a gap $\Delta$, but in BCS theory the gap is smaller, $\Delta \approx 1.7 \ T_c$.
 Observations on the cuprates indeed show a gap of order $\Delta = 4.0 \, T_c$  \cite{Gomes:2007}, showing that they are not weakly coupled superconductors.  In the early work on holographic superconductors, an attempt was made to measure this gap. Since the system was homogeneous, there was no Drude peak so people looked at $\lim_{\omega \rightarrow 0} {\rm Re}(\sigma)$. This was found to behave like $e^{-\Delta_\sigma/T}$ over a range of  low temperatures. In the probe limit (where the spacetime metric is fixed), one found a value close to what we find here, $\Delta_\sigma = 4.2\,T_c$ \cite{Hartnoll:2008vx,Horowitz:2008bn}, but when backreaction was included, $\Delta_\sigma$ became smaller and depended on the charge of the scalar field \cite{Hartnoll:2008kx}. With the lattice, we have a much better way to measure the superconducting gap and its intriguing that we get a realistic value even for a charge two condensate.\footnote{Curiously, in the probe limit at low temperature,  ${\rm Re}(\sigma)$ is strongly suppressed for  $\omega <\omega_g \approx 8T_c$, even if one changes $\Delta_\sigma$ by changing the mass of the scalar field and spacetime dimension \cite{Horowitz:2008bn}.}

 Another key difference from BCS is the presence of the constant $a$ in \eqn{rhon}. If we believe this extrapolation  then $\rho_n$ remains nonzero even at zero temperature indicating uncondensed spectral weight.\footnote{In a very different holographic realization of a two fluid model, it was found that $\rho_n$ remained nonzero at $T=0$ when the charge on the scalar field was small, but vanished when the scalar charge became larger \cite{Sonner:2010yx}.} Observations on the cuprates indeed show uncondensed spectral weight at $ T= 0$ \cite{Orenstein}. Of course the cuprates are d-wave superconductors in which the gap has nodes on the Fermi surface which probably contribute to this effect. It is  surprising to see   $\rho_n \ne 0$ at $T=0$ in an s-wave superconductor, as we have here. In quasiparticle language, it suggests that the normal component consists of two constituents, one of which is gapped and the other remains gapless. Of course quasiparticles may not be the right language for the strongly correlated dual system.
(A precursor to this was seen in the homogeneous case, where it was found that $\lim_{\omega \rightarrow 0} {\rm Re}(\sigma)$ was exponentially small but not zero at $T=0$ \cite{Horowitz:2009ij}.)

\begin{figure}
\centerline{
\includegraphics[width=.5\textwidth]{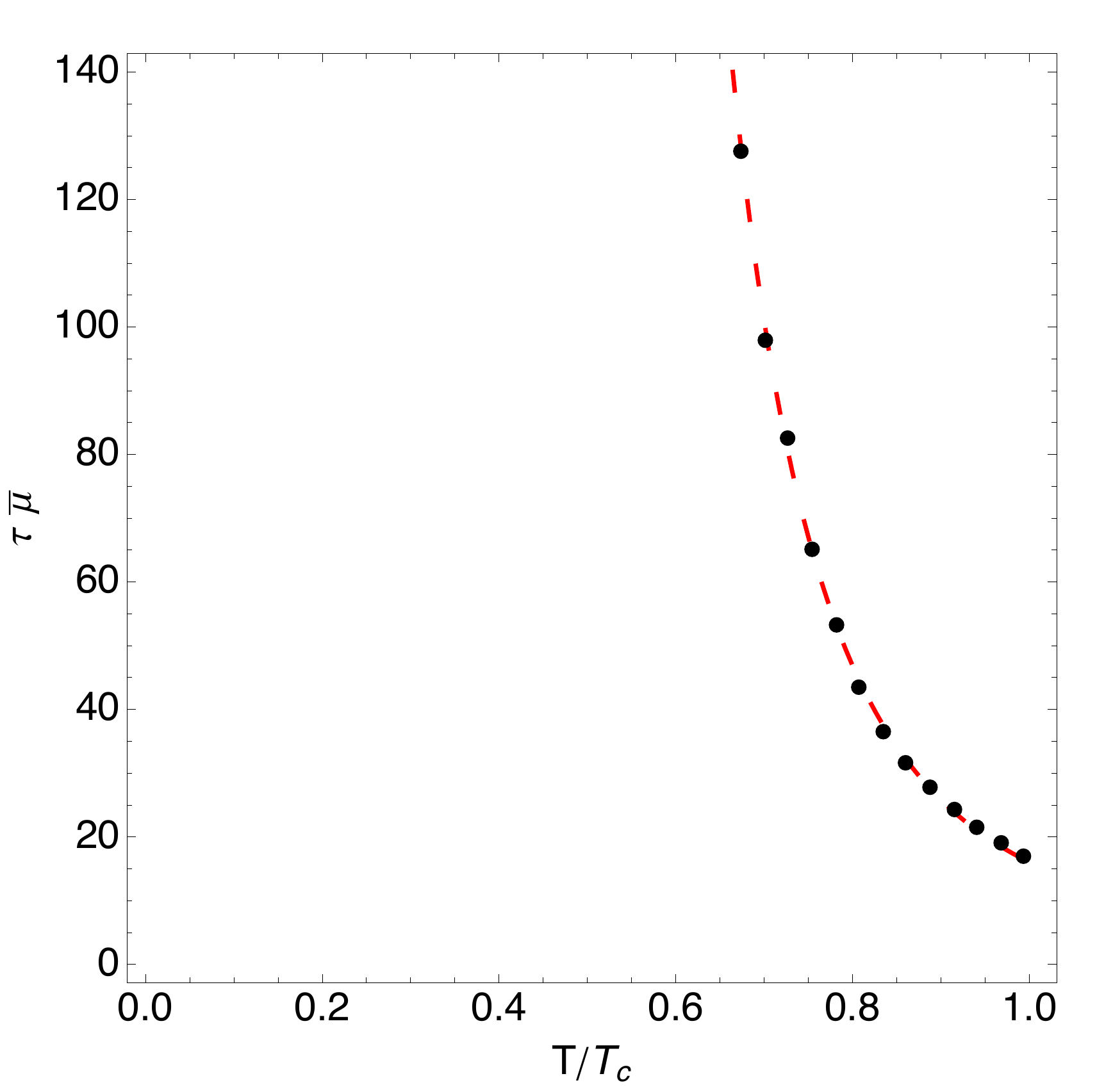}
}
\caption{The relaxation time as a function of temperature, extracted from the fit to \eqn{drudesuper}. The dashed curve is a fit to \eqn{gap}.}
\label{fig:tau}
\end{figure}

To check the extrapolation of $\rho_n$, one would like to compute $\rho_n$ at lower temperatures. This is very difficult because
the relaxation time $\tau$ rises rapidly as the temperature is lowered (see Fig.~\ref{fig:tau}). 
To see the Drude peak, one needs to probe frequencies  $\omega \tau \sim 1$. Since $\tau$ is rising so rapidly, the Drude peak is squeezed to very small frequencies which become difficult to resolve.
The dashed line in Fig.~\ref{fig:tau}  is a fit to 
\be\label{gap}
\tau = \tau_0\,e^{\tilde \Delta/T} \quad {\rm with}\quad \tilde  \Delta = 4.3 \ T_c\,.
\ee
It follows that the scattering rate, $1/\tau$, drops rapidly as $T$ is reduced below $T_c$. This is another observed feature of the cuprates \cite{Bonn}.  One often distinguishes clean and dirty limits of superconductors by comparing the scattering rate to the gap. It is easy to check that our holographic superconductor is in the clean limit: $1/\tau \ll 2\Delta$ for all $T < T_c$. 

It is natural to ask if $\rho_n$ is related to the charge inside the black hole. This appears not to be the case.
As we lower the temperature, the charge inside the horizon is reduced, and more of the charge is carried by the scalar field outside the horizon. However, in the limit $T\rightarrow 0$,  the charge inside  goes to zero as the horizon area vanishes, in contrast to \eqn{rhon}. 

\subsection{Power law}

\begin{figure}
\centerline{
\includegraphics[width=.6\textwidth]{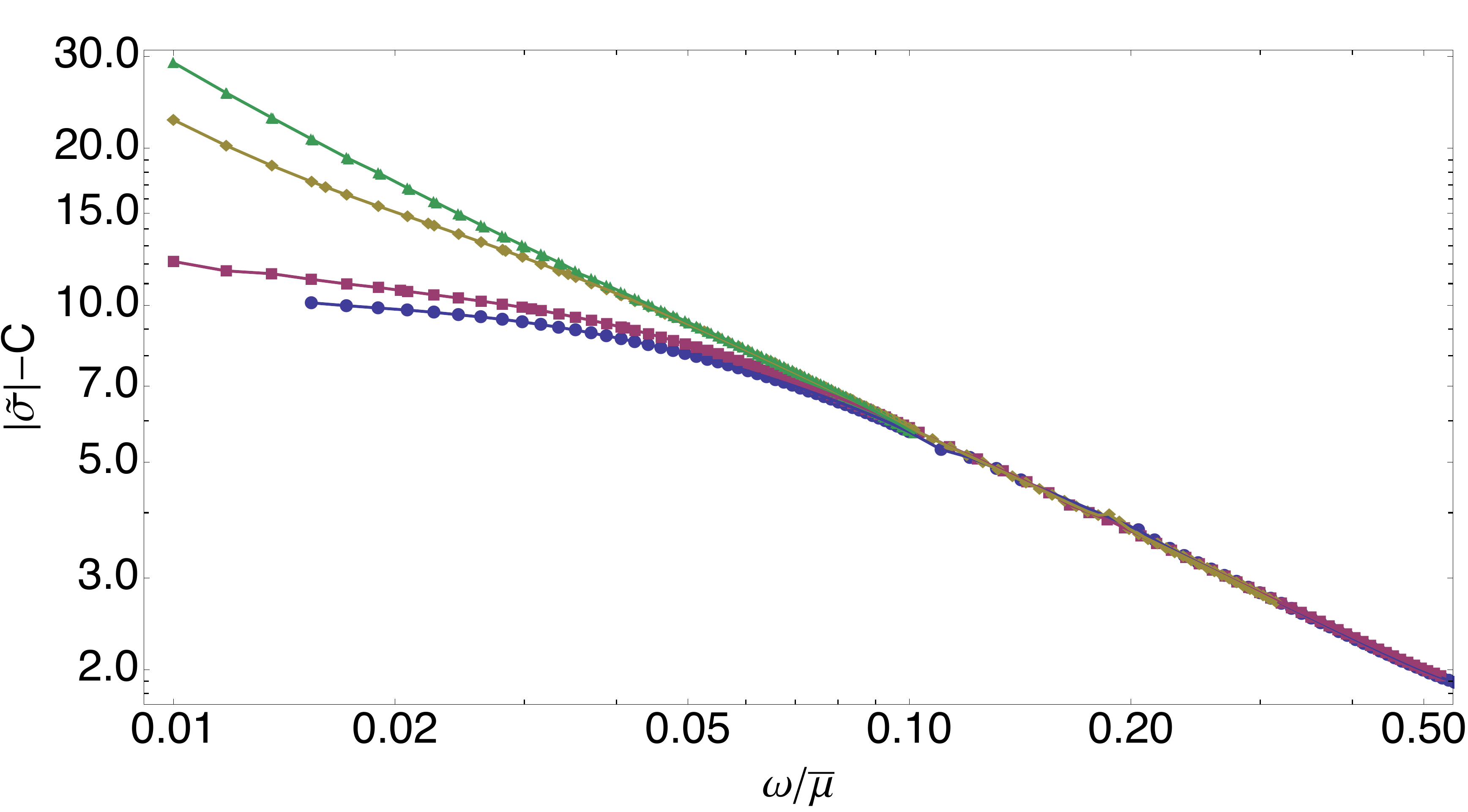}
}
\caption{The normal component of the conductivity on a log-log scale showing power law behavior with exponent $-2/3$. The pole in the imaginary part has been subtracted out. The lines correspond to $T/T_c = 1.0$ (blue circles), .97 (red squares), .86 (yellow diamonds),  and .70 (green triangles).}
\label{fig:powerlaw}
\end{figure}

 At slightly larger frequency, the absolute value of the conductivity of the normal component follows the same power law \eqn{power}  as was found in the normal phase above the critical temperature. This is shown in Fig. \ref{fig:powerlaw} where we have subtracted the pole in the imaginary part of $\sigma$ coming from the superfluid component, and plotted the absolute value of the result, $|\tilde \sigma| $, minus the off-set $C$ vs frequency on a log-log plot. The four lines correspond to the same temperatures as in Fig. \ref{fig:conductivity}. The fact that the lines are parallel implies that the exponent is the same. The fact that the lines lie on top of each other implies that the coefficient $B$ of the power law is again temperature independent. Recall that the lowest curve represents the normal phase. This clearly shows that the power law fall-off is completely unaffected by the superconducting phase transition.

\begin{figure}
\centerline{
\includegraphics[width=1.0\textwidth]{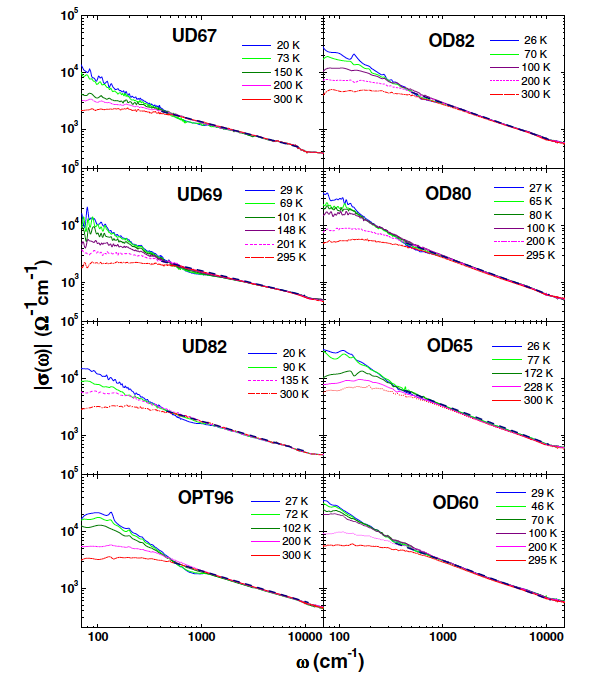}
}
\caption{A log-log plot of the optical conductivity in eight BSCCO samples  ranging from underdoped (UD) to overdoped (OD). The numbers following UD or OD in the plot labels are the critical temperatures. Each plot contains curves at several different temperatures both above and below $T_c$. The power-law fall-off is clearly unaffected by $T_c$. Plot is taken from \cite{hwang}.}
\label{fig:doping}
\end{figure}

Our results are in complete agreement with observations of the bismuth-based cuprates\footnote{We thank van der Marel for bringing this paper to our attention.} \cite{hwang}. Fig. \ref{fig:doping} shows measurements of $|\sigma(\omega)|$ (without the pole in the imaginary part) for $Bi_2Sr_2CaCu_2O_{8+\delta}$, commonly called BSCCO.   The figure includes eight separate log-log plots showing measurements on eight different samples ranging from underdoped with $T_c =  67K$ to the optimally doped with $T_c = 96K$ to overdoped with $T_c = 60$.  Each plot shows a variety of temperatures both above and below $T_c$. All show power-law behavior at intermediate frequencies which does not change as the temperature is reduced below $T_c$.  The measured exponent is $-2/3$ for the optimally doped and overdoped samples but increases to about $-1/2$ in the most underdoped samples. The agreement between Fig. \ref{fig:powerlaw} and Fig. \ref{fig:doping} is striking, and we do not understand why our simple gravity model is able to reproduce this key feature of BSCCO so well. Of course one important difference is that the experimental measurements do not show an off-set $C$. Another difference is that the power law in BSCCO extends up to a temperature independent frequency (the lower cut-off is temperature dependent). In our calculations the power law always extends roughly between $2< \omega \tau < 8$ which is exactly the same range as was found in the normal phase $T > T_c$ \cite{Horowitz:2012ky}. Since $\tau$ is temperature dependent, both the upper and lower cut-offs are temperature dependent.

\begin{figure}
\centerline{
\includegraphics[width=.8\textwidth]{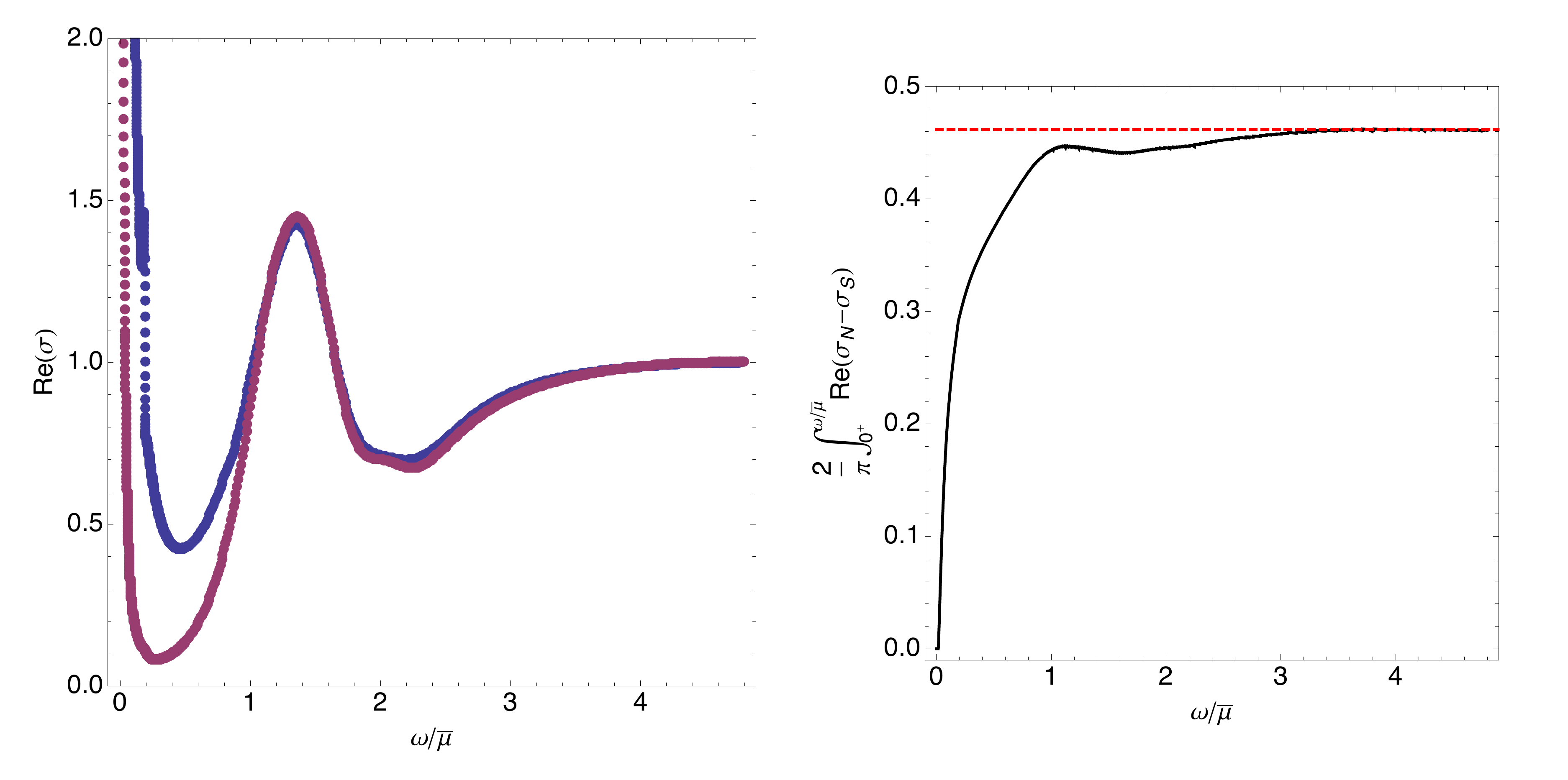}
}
\caption{\emph{Left panel}: The conductivity in the normal phase $T=T_c$ (upper blue curve) and superconducting phase $T = .71\ T_c$ (lower red curve). \emph{Right panel}: The red dashed line is the superfluid density obtained from the pole in the imaginary part of the conductivity. The black line is  $(2/\pi) \int_{0^+}^{\omega/\bar\mu} d\tilde \omega {\rm Re}[ \sigma_N(\tilde \omega) - \sigma_S(\tilde \omega)] $. This integral approaches the red line at large $\omega$ showing that the FGT sum rule \eqn{sumrule} is satisfied.}
\label{fig:comparison}
\end{figure}

The phase of the complex conductivity computed from the gravitational dual is roughly constant over the range of frequencies where the magnitude follows the power law. However, this phase is temperature dependent and varies between $60^o$ and $80^o$. The data on BSCCO shows a temperature independent phase of $60^o$. This difference is likely connected with the constant off-set in our power law. Without the offset, scale invariance, causality and time reversal symmetry require $\sigma(\omega) \propto (-i\omega)^\alpha$ \cite{vandermarel}, so the phase is related to the exponent of the power law.

\subsection{Sum rule}

The Ferrell-Glover-Tinkham (FGT) sum rule states that the reduction in the spectral weight  when $T<T_c$ is taken up by the superfluid density $\rho_s$:
\be\label{sumrule}
\int_{0^+}^\infty d\omega {\rm Re}[ \sigma_N(\omega) - \sigma_S(\omega)] =\frac{\pi}{2}\rho_s 
\ee
where $\sigma_N$ is the conductivity in the normal phase and $\sigma_S$ is the conductivity in the superconducting phase. We now ask whether this is satisfied in our gravitational model\footnote{See \cite{Gulotta:2010cu,WitczakKrempa:2012gn,WitczakKrempa:2013ht} for  general discussions of sum rules from gravity for translationally invariant backgrounds.}. The left panel of Fig.~\ref{fig:comparison} shows Re$[\sigma(\omega)]$ over a wide range of frequency in both the normal phase $T=T_c$ and the superconducting phase $T= .71\ T_c$. The right panel of Fig.~\ref{fig:comparison}  shows that $(2/\pi) \int_{0^+}^{\omega/\bar\mu} d\tilde\omega {\rm Re}[ \sigma_N(\tilde\omega) - \sigma_S(\tilde\omega)] $ indeed approaches  the superfluid density at large $\omega$. At the end of the integration, they differ by less than $0.6\%$. This is not only a confirmation that our holographic superconductor satisfies the FGT sum rule, but also a strong test of the numerics. The two curves on the left panel of Fig.~\ref{fig:comparison} were computed using different codes, and the fact that an integral of their difference gives precisely the expected answer is confirmation of the accuracy of the numerical results. 

Note that the two curves on the left panel of Fig.~\ref{fig:comparison} differ over a range of frequencies of order $\bar \mu$. In conventional superconductors, Re$[\sigma(\omega)]$ is reduced only over a much smaller range of frequency, so it is common to cut off  the integral in \eqn{sumrule} at a convenient low frequency $\omega_0$. However, when this is done in the cuprates, one finds that the integral underestimates the superfluid density \cite{hwang}. In other words, some of the spectral weight represented by $\rho_s$ is missing from the low frequency part of Re$[\sigma(\omega)]$. One must indeed include frequencies of order the chemical potential to recover the sum rule, just as we have seen here \cite{Boris}.
%
%
%\begin{figure}
%\centerline{
%\includegraphics[width=.5\textwidth]{sum_rule.pdf}
%}
%\caption{The red dashed line is the superfluid density obtained from the pole in the imaginary part of the conductivity. The black line is  $(2/\pi) \int_{0^+}^{\omega/\bar\mu} d\tilde \omega {\rm Re}[ \sigma_N(\tilde \omega) - \sigma_S(\tilde \omega)] $. This integral approaches the red line at large $\omega$ showing that the FGT sum rule \eqn{sumrule} is satisfied.}
%\label{fig:sum_rule}
%\end{figure}

\subsection{Resonance}

The resonance in the conductivity at $\omega/\bar\mu \approx 1.5$ can be understood on the gravity side as arising from a quasinormal mode oscillation of the black hole, just like the example in \cite{Horowitz:2012gs}. As discussed in \cite{Son:2002sd} quasinormal modes correspond to poles in retarded Green's functions. We can determine the quasinormal mode frequency by fitting the conductivity to 
\be\label{qnm}
\sigma(\omega) = \frac{G^R(\omega)}{i\omega}= \frac{1}{i\omega} \frac{a+b(\omega - \omega_{0})}{\omega - \omega_{0}}\,.
\ee
For $T= T_c$, one finds the resonance is very well fit by this expression with  $\omega_0/\bar\mu = 1.48 -0.42i$.  This quasinormal mode changes very little when the temperature drops to $.7\ T_c$.

 \section{Discussion}
 
We have seen that a simple holographic model of a superconductor reproduces quantitative features of BSCCO including an intermediate frequency power law with exponent $-2/3$ which is unaffected by the superconducting phase transition, and a superconducting gap $\Delta \approx 4T_c$. In addition, it reproduces many qualitative features of the cuprates including a rapidly decreasing scattering rate below $T_c$,  uncondensed spectral weight at $T=0$, and a superconductivity induced transfer of spectral weight involving energies of order the chemical potential.
 
 We find it remarkable that so much of the phenomenology of the cuprates can be reproduced by such a simple gravity model. 
 Since gravity coupled to a Maxwell field and charged scalar are part of the low energy limit of string theory with many different compactifications, one can perhaps view it as the universal part of the theory. In this respect, its dual description might be called the ``standard model" of strongly correlated systems. One could start with this and then add other fields to reproduce observed fine scale structure of particular materials. Perhaps the main drawback of such an approach is that our current model  describes an $s$-wave superconductor. We still do not have a satisfactory description of a $d$-wave holographic superconductor. See \cite{Benini:2010pr}  for a description of some of the difficulties.
 
 We have presented results for the optical conductivity in the superconducting region for only one value of the lattice amplitude $A_0 =2$ and one value of the wavenumber $k_0 =2$ (with $\bar\mu =1$). We believe that the $-2/3$ exponent in the power law is independent of these choices since it is unchanged when one enters the superconducting regime, and above $T_c$ it has been shown to be very robust. We have done preliminary calculations with $k_0 =1 $ which indicate that the behavior of the normal component density $\rho_n(T)$ is also independent of these choices. In particular, although $T_c$ increases,  the gap remains $\Delta = 4T_c$ and $\rho_n$ approaches the same nonzero value as $T\rightarrow 0$. 
 The behavior of the relaxation time $\tau$ is qualitatively the same, but there is an important difference.  Although one again finds $\tau = \tau_0 e^{\tilde \Delta/T}$, the value of $\tilde \Delta$ is half what it was for $k_0 =2$. In other words, it appears that $1/\tau \propto e^{-ak_0/T}$. This is the behavior predicted in \cite{Hartnoll:2012rj} where the scattering rate was related to the density-density correlation function at $\omega = 0$ and $k = k_0$. Since Poincar\'e invariance appears to be restored in the zero temperature infrared geometry \eqn{nhmetric}, low energy physical states should have $\omega \sim k$ so states with  $\omega = 0$ and $k = k_0$ would be exponentially suppressed.
  
 The fact that we still see -2/3 power law even in the superconducting phase, makes it clear that this power law has nothing to do with the $AdS_2\times R^2$
 near horizon geometry of the zero temperature Reissner-Nordstr\"om solution. As we have discussed, with the charged scalar present, the zero temperature solution has a singular horizon.
 
 Homes has discovered a remarkable relation between the low temperature superfluid density $\rho_s$, the critical temperature $T_c$ and the DC conductivity just above the critical temperature $\sigma_{dc}$ \cite{Homes}. He found that for a wide range of high temperature superconductors
\be
\rho_s \approx 35\,\sigma_{dc}\,T_c\label{homes}\,.
\ee
This relation appears to be universal. It holds for the conductivity perpendicular to the $CuO_2$ planes in the cuprates as well as the conductivity in the plane, and also holds for the iron pnictides. 
Unfortunately, it is easy to see that Homes' law cannot hold for the class of holographic superconductors we have discussed here where the critical temperature depends on the lattice amplitude.\footnote{An earlier unsuccessful attempt  to find a holographic realization of Homes law was made in \cite{Erdmenger:2012ik}.} The reason is that as $A_0 \rightarrow 0$, $\sigma_{dc}$ diverges since the system becomes translationally invariant. However, both $\rho_s$ and $T_c$ approach finite limits given by the original homogeneous holographic superconductor. So \eqn{homes} cannot hold. To reproduce \eqn{homes} one would have to fix a large $A_0$ and vary $T_c$ by other means, perhaps by adding a double trace perturbation \cite{Faulkner:2010gj}.

\vskip 1cm
\centerline{\bf Acknowledgements}
\vskip 1cm

It is a pleasure to thank D. Tong for collaboration at an early stage of this project. We also thank N. Iqbal and the participants of the Simons  Symposium on  Quantum Entanglement: From Quantum Matter to String Theory, especially A. Millis, S. Sachdev, and J. Zaanen for discussions.
This work was supported in part by the National Science Foundation under Grant No. PHY12-05500

\appendix
\section{The first law for black holes with varying chemical potential}

In this appendix, we derive the first law for electrically charged black hole solutions to \eqn{eq:action} with nonconstant chemical potential, $\mu(x,y)$. For convenience we will imagine that the $x$ and $y$ directions are periodically identified so the black holes have finite horizon area and mass. We follow the approach of Sudarsky and Wald \cite{Sudarsky:1992ty}. As usual for a diffeomorphism invariant theory, the Hamiltonian is a linear combination of the constraints plus surface terms: 
\be\label{hamiltonian}
H = \int_\Sigma [N^a C_a + N^a A_a (D_i E^i)]  +  {\rm surface \ terms}
\ee
where $N^a$ is the lapse-shift vector and $C_a$ are the usual Hamiltonian and momentum constraints of general relativity. The surface terms are determined by the requirement that the variations of $H$ with respect to the canonical variables $q_{ij}, p^{ij}, A_i, E^i$  are well defined. In addition to the usual gravitational surface terms, there is an additional term coming from the Gauss law constraint:
\be\label{em}
\oint_{\partial \Sigma} (N^a A_a) E_i dS^i 
\ee

Given a static, electrically charged black hole, we choose $\Sigma$ to be a constant $t$ surface which starts at the bifurcation surface on the horizon and ends at infinity. We also choose $N^a = (\partial/\partial t)^a$. Now under a perturbation of the black hole 
\be
\delta H = \int \delta q_{ij} \partial H/\partial q_{ij} + \cdots =0
\ee
 since the partial derivatives of the Hamiltonian just yield the time derivatives of the canonical variables in the background which vanish since the background is static.
 On the other hand, if the perturbation solves the linearized field equations, it certainly solves the linearized constraints,  so from \eqn{hamiltonian} $\delta H$ reduces to a sum of the variation of the surface terms. The gravitational surface terms yield the usual $\delta M - (\kappa/8\pi) \delta A$. Since $\partial/\partial t = 0$ at the bifurcation surface, the electromagnetic surface term \eqn{em} only has a contribution from infinity which is simply $\int \mu(x,y) \rho(x,y) dx dy$. Since $\mu(x,y)$ is fixed by our boundary conditions, the variation of this surface term is $\int \mu(x,y)\delta \rho(x,y) dx dy$. We thus obtain the first law:
 \be
 \delta M =\frac{\kappa}{8\pi} \delta A + \int \mu(x,y) \delta \rho(x,y) dx dy
 \ee
 In the homogeneous case with constant $\mu$, this clearly reduces to the familiar form
  \be
 \delta M =\frac{\kappa}{8\pi} \delta A + \mu\delta Q\ .
 \ee

%%%%%%%%%%%%%%%%%%%%%%%%%%%%%%%%%%%%%%%%%%%%%%%%%%%%%%%%%%%%%%%%%
% BIBLIOGRAPHY
%%%%%%%%%%%%%%%%%%%%%%%%%%%%%%%%%%%%%%%%%%%%%%%%%%%%%%%%%%%%%%%%%

\bibliographystyle{JHEP}
\bibliography{superlattice.bib}

\end{document}